\documentclass[reprint,aps,prb,amsmath,amssymb,groupedaddress]{revtex4-2}
\usepackage{graphicx}
\usepackage{dcolumn}
\usepackage{bm}
\usepackage{color} 
\usepackage[tight]{subfigure}
\usepackage{hyperref}

\newcommand{\ket}[1]{\left|{}#1 \right>}
\newcommand{\bra}[1]{\left<{}#1 \right|}
\newcommand{\expect}[1]{\left<{}#1\right>}

\newcommand{\rmi}{\mathrm{i}}
\newcommand{\rme}{\mathrm{e}}

\begin{document}
	
	\title{Quantum phase transitions in the triangular coupled-top model}
	\author{Liwei Duan$^{1}$}
	\email{duanlw@gmail.com}
	\author{Yan-Zhi Wang$^{2}$}
	\author{Qing-Hu Chen$^{3,4}$}%
	\email{qhchen@zju.edu.cn}
	\address{
		$^{1}$ Department of Physics, Zhejiang Normal University, Jinhua 321004, China \\
		$^{2}$ School of Physics and  Electronic Information, Anhui Normal University, Wuhu 241002, China \\
		$^{3}$ School of Physics, Zhejiang University, Hangzhou 310027, China \\
		$^{4}$ Collaborative Innovation Center of Advanced Microstructures, Nanjing University, Nanjing 210093, China
	}

	\date{\today}
	
	\begin{abstract}
		We study the coupled-top model with three large spins located on a triangle. Depending on the coupling strength, there exist three phases: disordered paramagnetic phase, ferromagnetic phase,
		and frustrated antiferromagnetic phase, which can be distinguished by the mean-field approach. The paramagnetic-ferromagnetic phase transition is accompanied by the breaking of the global $Z_2$ symmetry, whereas the paramagnetic-antiferromagnetic phase transition is accompanied by the breaking of both the global $Z_2$ symmetry and the translational symmetry. Exact analytical results of higher-order quantum effects beyond the mean-field contribution, such as the excitation energy, quantum fluctuation, and von Neumann entropy, can be achieved by the Holstein-Primakoff transformation and symplectic transformation in the thermodynamic limit. Near the quantum critical point, the energy gap closes, along with the divergence of the quantum fluctuation in certain quadrature and von Neumann entropy.
		Particular attention should be paid to the antiferromagnetic phase, where geometric frustration takes effect. The critical behaviors in the antiferromagnetic phase are quite different from those in the paramagnetic and ferromagnetic phases, which highlight the importance of geometric frustration. The triangular coupled-top model provides a simple and feasible platform to study the quantum phase transition and the novel critical behaviors induced by geometric frustration.
	\end{abstract}
	
	\maketitle
	
	\section{Introduction}
	
	Quantum phase transitions are usually accompanied by a qualitative change in the nature of the ground state when the control parameter of a system passes through the quantum critical point. In contrast to its classical counterpart driven by thermal fluctuations, the quantum phase transition occurs at zero temperature due to the quantum fluctuations, where the non-commuting terms in the system Hamiltonian play a dominant role \cite{sachdev_2011}. The geometric frustration, which prohibits simultaneously satisfying every pairwise interaction due to local geometric constraints, also influences the quantum phase transition significantly \cite{PhysRev.79.357,Toulouse1977,Balents2010}. Frustration is closely associated with exotic materials such as quantum spin liquid and spin ice \cite{Balents2010,RevModPhys.89.025003,doi:10.1126/science.1064761}, as well as complex networks \cite{RevModPhys.80.1275} and protein folding \cite{doi:10.1073/pnas.84.21.7524}, etc. The unconventional quantum critical phenomena and  qualitatively new states of matter induced by the frustration are still an active area of research \cite{Lee2002-cy,doi:10.1126/sciadv.1500001,PhysRevLett.127.063602,PhysRevLett.129.183602,PhysRevLett.128.163601,doi:10.1146/annurev.ms.24.080194.002321,doi:10.1139/p97-007}.
	
	As a paradigmatic system to study quantum phase transitions, the transverse-field Ising model (TFIM) has drawn persistent attention for over half a century \cite{DEGENNES1963132,BROUT1966507,PFEUTY197079,Stinchcombe_1973,sachdev_2011,Suzuki2012-ro,PhysRevLett.95.245701}. Besides, the TFIM with antiferromagnetic interaction on a two-dimensional triangular lattice offers itself as an ideal platform to study the frustrated magnetic behaviors \cite{PhysRevB.63.224401,Kim2010-au}. A natural generalization, named coupled-top model, is to replace the widely used spin-$\frac{1}{2}$s in the TFIM with large spins \cite{PhysRevA.71.042303,PhysRevE.102.020101,PhysRevE.104.024217,PhysRevE.104.034119,PhysRevE.105.014130}.
	The coupled-top model shares some similarities with the Dicke model, a fundamental model describing the interaction between light and matter \cite{PhysRevE.67.066203,PhysRevLett.90.044101}. (i) There exist the quantum phase transition and spontaneous symmetry breaking in both models: the global $Z_2$ symmetry is spontaneously broken with increasing coupling strength, and the von Neumann entropy diverges near the quantum critical point \cite{PhysRevA.71.042303,PhysRevLett.92.073602}. (ii) The level statistics changes from Poissonian distribution to Wigner-Dyson distribution, which indicates a transition from quasi-integrability to quantum chaos \cite{PhysRevE.102.020101,PhysRevE.67.066203}. (iii) The density of states shows a singular behavior that corresponds to the excited-state quantum phase transitions \cite{PhysRevE.88.032133,PhysRevE.104.034119}. Recently, the light-matter interacting systems in a triangular structure have intrigued an enormous amount of attention due to the exotic behaviors, such as chiral superradiant phase \cite{PhysRevLett.127.063602,PhysRevLett.129.183602}, frustrated superradiant phase and novel critical exponent \cite{PhysRevLett.128.163601}. We expect that the triangular coupled-top model also possesses rich quantum phases and unconventional quantum critical phenomena. 
	
	In this paper, we study the quantum phase transition in the triangular coupled-top model. Three large spins are located on a triangle. Each pairwise interaction can be ferromagnetic or antiferromagnetic. First, we introduce the mean-field approach which has been widely employed to study large spin systems, especially in the thermodynamic limit \cite{Lieb1973-sw,PhysRevB.71.224420,PhysRevA.100.063820}. The mean-field approach is equivalent to using a product state constructed by SU(2) coherent states  as the trial wavefunction. The ground state is achieved by minimizing the corresponding energy expectation value. With the mean-field approach, we confirm the existence of the quantum phase transition by the nonzero order parameter, which indicates the breaking of global $Z_2$ symmetry. Two critical coupling strengths separate the coupled-top model into three phases: disordered paramagnetic phase, ferromagnetic phase, and frustrated antiferromagnetic phase. Then, we introduce the Holstein-Primakoff transformation and symplectic transformation in order to capture the quantum fluctuation and quantum entanglement which are ignored by the mean-field approach. The Holstein-Primakoff transformation maps the collective spin operators into bosonic ones, which leads to an effective quadratic Hamiltonian in the thermodynamic limit. The covariance matrix of the effective quadratic Hamiltonian can be obtained by the symplectic transformation, with which we can analyze the behaviors of the quantum fluctuation and von Neumann entropy. We find that geometric frustration comes into play in the antiferromagnetic phase, which leads to novel critical behaviors.
	
	The paper is structured as follows. In Sec. \ref{sec:model}, we introduce the triangular coupled-top model and the symmetry it possesses. In Sec. \ref{sec:MF}, the quantum phase transition is confirmed by the mean-field approach. Exact analytical results of higher-order quantum effects beyond the mean-field contribution, as well as the corresponding critical behaviors, are given in Sec. \ref{sec:MF_beyond}. A brief summary is given in Sec. \ref{sec:summary}.

	\section{Triangular Coupled-top model} \label{sec:model}
	
	\begin{figure*}
		\includegraphics[scale=0.25]{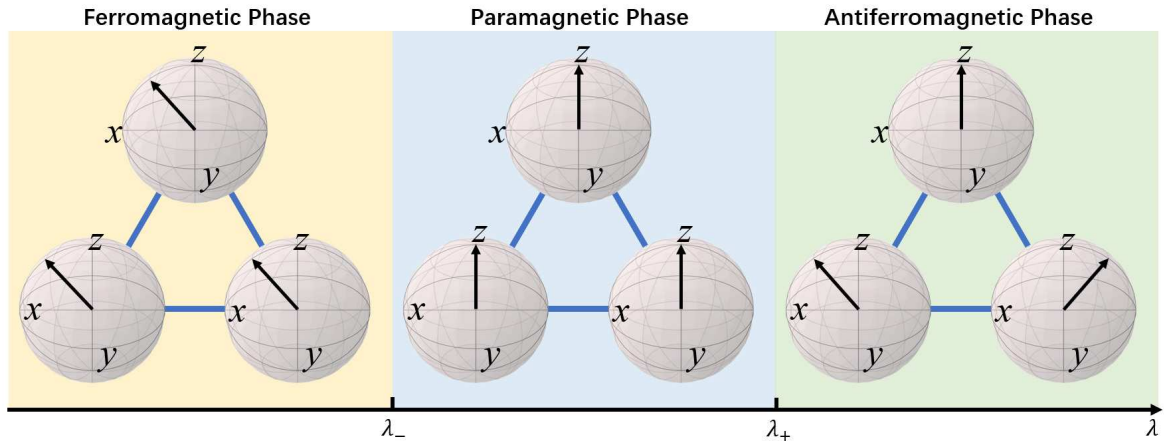}
		\caption{\label{fig:bloch} Schematic illustration of different phases in the coupled-top model: Three large spins are located on a triangle. Their directions are specified by the black arrows on the Bloch sphere. $\lambda_- < \lambda < \lambda_+$ corresponds to the disordered paramagnetic phase with all spins aligning in parallel to the external field in $z$ axis. $\lambda < \lambda_-$ corresponds to the ferromagnetic phase where all spins tend to align in parallel along the $x$ axis. $\lambda > \lambda_+$ corresponds to the frustrated antiferromagnetic phase. Two of the large spins tend to align antiparallel along the $x$ axis, whereas the third one is frozen in $z$ axis. $\lambda_+ = 1$ and $\lambda_- = -\frac{1}{2}$.}
	\end{figure*}
	
	The coupled-top model can be regarded as the generalization of the TFIM with spin-$\frac{1}{2}$s replaced by large spins \cite{PhysRevA.71.042303,PhysRevE.102.020101,PhysRevE.104.024217,PhysRevE.104.034119,PhysRevE.105.014130}. Previous studies mainly focused on two coupled large spins where the influences of the geometric structure are negligible. In order to introduce the geometric frustration, we study the triangular coupled-top model with $N=3$ large spins, whose Hamiltonian can be written as
	\begin{eqnarray}
		\hat{H}_{\text{CT}} = \sum_{i=1}^N \left(-\epsilon \hat{J}_i^z + \frac{\chi}{J} \hat{J}_i^x \hat{J}_{i + 1}^x\right) . 
	\end{eqnarray}
	Here $\epsilon$ is the strength of the external field, and $\chi$ is the coupling strength. 
	$\hat{\mathbf{J}}_i = \left(\hat{J}_i^x, \hat{J}_i^y, \hat{J}_i^z\right)$ are the collective spin operators which describe the $i$th large spin. The eigenvalues of $\hat{\mathbf{J}}_i^2$ are $J(J+1)$.
	The periodic boundary condition is performed which leads to $\hat{J}_{N+1}^r = \hat{J}_1^r$ with $r=x,y,z$.
	For simplicity, we introduce the dimensionless form as follows,
	\begin{eqnarray}
		\hat{H} = \hat{H}_{\text{CT}} / \epsilon = \sum_{i=1}^N \left(- \hat{J}_i^z + \frac{\lambda}{J} \hat{J}_i^x \hat{J}_{i + 1}^x\right) , \label{eq:H}
	\end{eqnarray}
	with the dimensionless coupling strength $\lambda = \chi / \epsilon$.
	Hamiltonian (\ref{eq:H}) commutes with the parity operator $\hat{\Pi}$, with 
	\begin{eqnarray}
		\hat{\Pi} = \exp \left[\rmi \pi \sum_{i=1}^{N} \left(\hat{J}_i^z + J\right)\right] ,
	\end{eqnarray}
	which indicates a global $Z_2$ symmetry. 
	The expectation values satisfy $J_i^x = 0$ if the $Z_2$ symmetry is preserved, whereas  $J_i^x \neq 0$ if the $Z_2$ symmetry is broken. Therefore, $J_i^x$ plays a role of the order parameter.
	What's more, Hamiltonian (\ref{eq:H}) admits a translational symmetry, which indicates that $J_i^x$ doesn't depend on the site index $i$ in the symmetric phase.
	
	\section{Mean-field approach} \label{sec:MF}
	Let us first discuss the coupled-top model in the mean-field picture, which ignores the correlations between different large spins. We can construct a mean-field ansatz for the ground state by employing a product state \cite{PhysRevB.71.224420,PhysRevA.85.043821},
	\begin{eqnarray}
		\ket{\psi} = \ket{\theta_1, \phi_1} \otimes \ket{\theta_2, \phi_2} \otimes \ket{\theta_3, \phi_3} ,
	\end{eqnarray}
	with $\ket{\theta_{i}, \phi_i}$ ($0 \le \theta_i \le \pi$, $0 \le \phi_i < 2 \pi$) the SU(2) coherent state \cite{Perelomov2012-qw,PhysRevA.105.042215} defined as
	\begin{eqnarray}
		\ket{\theta_i, \phi_i} = \exp \left[\frac{\theta_i}{2} \left(\rme^{\rmi \phi_i} \hat{J}_i^- - \rme^{-\rmi \phi_i} \hat{J}_i^+\right)\right] \ket{J, J} .
	\end{eqnarray}
	The corresponding expectation values of the collective spin operators can be written as
	\begin{eqnarray}
		\left(J_i^x, J_i^y, J_i^z\right) &=& \bra{\theta_i, \phi_i} \left(\hat{J}_i^x, \hat{J}_i^y, \hat{J}_i^z\right) \ket{\theta_i, \phi_i} \\
		&=& J \left(\sin \theta_i \cos \phi_i, \sin \theta_i \sin \phi_i, \cos \theta_i\right), \nonumber
	\end{eqnarray}
	which corresponds to a point on the Bloch sphere. Note that the external field creates an energetic bias for one $z$-axis spin direction over the other one, whereas the coupling strength leads to the alignment or anti-alignment of spin projections along the $x$ axis. Therefore, the large spins tend to be located on the $x$-$z$ plane which have $\phi_i = 0$ or $\pi$. Without loss of generality, we can introduce $\tilde{\theta}_i$,
	\begin{eqnarray}
		\tilde{\theta}_i = \left\{\begin{array}{c}
			\theta_i, \text{ for } \phi_i = 0, \\
			-\theta_i, \text{ for } \phi_i = \pi.
		\end{array}\right.
	\end{eqnarray}
	The expectation values of the collective spin operators can be simplified as
	\begin{eqnarray}
		\left(J_i^x, J_i^y, J_i^z\right) &=& J \left(\sin \tilde{\theta}_i, 0, \cos \tilde{\theta}_i\right), \nonumber
	\end{eqnarray}
	with which we can achieve the energy expectation value given by
	\begin{eqnarray} \label{eq:E_GS}
		E &=& \frac{1}{J} \bra{\psi} \hat{H} \ket{\psi} \\
		&=& \sum_{i=1}^N - \cos \tilde{\theta}_i + \lambda \sin \tilde{\theta}_i \sin \tilde{\theta}_{i + 1} .\nonumber
	\end{eqnarray}
	The ground state is determined by minimizing the energy expectation value $E$ with respect to $\left\{\tilde{\theta}_i \right\}$, namely,	
	\begin{eqnarray} \label{eq:E_min}
		\frac{\partial E}{\partial \tilde{\theta}_i} &=&  \sin \tilde{\theta}_i + \lambda \cos \tilde{\theta}_i \left(\sin \tilde{\theta}_{i + 1} + \sin \tilde{\theta}_{i - 1}\right) = 0.
	\end{eqnarray}
	By carefully analyzing the energy expectation value, we find two critical coupling strengths $\lambda_- = -\frac{1}{2}$ and $\lambda_+ = 1$, which separate the coupled-top model into three phases: disordered paramagnetic phase, ferromagnetic phase, and frustrated antiferromagnetic phase, as shown in Fig. \ref{fig:bloch}.
	
	\begin{figure}
		\includegraphics[scale=0.95]{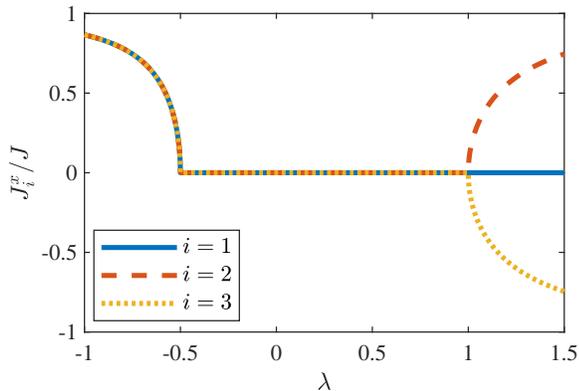}
		\caption{\label{fig:jx} The order parameters $J_i^x$ as a function of the dimensionless coupling strength $\lambda$. The blue solid line, red dashed line and yellow dotted line correspond to $J_1^x$, $J_2^x$ and $J_3^x$ respectively.}
	\end{figure}
	
	\subsection{Disordered paramagnetic phase} \label{sec:MF_Disorder}
	
	When the coupling strength is weak, the external field plays a dominant role. The large spins tend to align in parallel to the external field which leads to a disordered paramagnetic phase. For $\lambda_- < \lambda < \lambda_+$, the ground state is nondegenerate with $\tilde{\theta}_i = 0$ and $E_{\min} = -3 \epsilon$, as given in Eq. (\ref{eq:E_GS}). The coupled-top model preserves the global $Z_2$ symmetry and the translational symmetry,  with the order parameter $J_i^x = 0$ as shown in Fig. \ref{fig:jx}. 
	
	\subsection{Ordered phase}
	
	When the coupling strength is strong, the large spins prefer the alignment or anti-alignment of spin projections along the $x$ axis depending on the sign of $\lambda$.
	
	\subsubsection{Ferromagnetic phase}
	
	For $\lambda < \lambda_- = -\frac{1}{2}$, three large spins prefer to align in parallel along the $x$ axis to decrease the energy expectation value $E$ [Eq. (\ref{eq:E_GS})]. The ground state corresponds to 
	\begin{eqnarray} \label{eq:theta_ferro}
		\tilde{\theta}_i = \pm \arccos\left(-\frac{1}{2\lambda} \right),
	\end{eqnarray}
	and it is two-fold degenerate with the order parameters satisfying $J_i^x = \pm \sqrt{1 - \frac{1}{4\lambda^2}}$ respectively. We only show the positive branch of $J_i^x$ in Fig. \ref{fig:jx}. The critical exponent associated with the order parameter is $\beta = \frac{1}{2}$ due to $J_i^x \propto \left(\lambda_- - \lambda\right)^{1/2}$ near the quantum critical point ($\lambda \rightarrow \lambda_-$). The global $Z_2$ symmetry is broken in the ferromagnetic phase, whereas the translational symmetry is preserved.
	
	\subsubsection{Frustrated antiferromagnetic phase}
	
	For $\lambda > \lambda_+ = 1$, the energy expectation value is minimized when each spin is aligned opposite to its neighbors. However, once two of the large spins align antiparallel, the third one is frustrated and cannot align antiparallel with both of them simultaneously  due to the triangular arrangement. In the frustrated antiferromagnetic phase, Both the global $Z_2$ symmetry and the translational symmetry are broken, and the ground state is six-fold degenerate with
	\begin{eqnarray}
		\tilde{\theta}_i = 0, \quad \tilde{\theta}_{i + 1} = - \tilde{\theta}_{i - 1} = \pm \arccos \left(\frac{1}{\lambda}\right) .
	\end{eqnarray}
	The breaking of $Z_2$ symmetry leads to nonzero $J_i^x$, while the breaking of translational symmetry leads to dependence of $J_i^x$ on the site index $i$.
	For clarity, we only present one of the six degenerate ground-state solutions, which corresponds to $J_1^x = 0$ and $J_2^x = -J_3^x = \sqrt{1 - \frac{1}{\lambda^2}}$, as depicted in Fig. \ref{fig:jx}. The order parameter satisfies $J_2^x = -J_3^x \propto \left(\lambda - \lambda_+\right)^{1/2}$ near the quantum critical point ($\lambda \rightarrow \lambda_+$), which also yields the critical exponent associated with the order parameter $\beta = \frac{1}{2}$.
	
	\section{Beyond mean-field approach} \label{sec:MF_beyond}
	
	In the thermodynamic limit ($J \rightarrow + \infty$), the mean-field approach is able to describe the ground-state energy and order parameters very well \cite{Lieb1973-sw,PhysRevA.100.063820}, from which we can distinguish different phases. However, one needs to go beyond the mean-field approach to study the correlations between different large spins, as well as the quantum fluctuations. 
	To separate the mean-field contribution and the higher-order terms, we first perform a unitary transformation \cite{PhysRevB.71.224420,PhysRevLett.128.163601,PhysRevA.104.052423}, which leads to
	\begin{widetext}
		\begin{eqnarray} \label{eq:H_R}
			\hat{\bar{H}} &=& \hat{U}^{\dagger} \hat{H} \hat{U} = \sum_{i=1}^N - \left(\cos \tilde{\theta}_i \hat{J}_i^z + \sin \tilde{\theta}_i \hat{J}_i^x\right) +  \frac{\lambda}{J} \left(\cos \tilde{\theta}_i \hat{J}_i^x - \sin \tilde{\theta}_i \hat{J}_i^z\right) \left(\cos \tilde{\theta}_{i + 1} \hat{J}_{i + 1}^x - \sin \tilde{\theta}_{i + 1} \hat{J}_{i + 1}^z\right) , 
		\end{eqnarray}
	\end{widetext}
	with 
	\begin{eqnarray}
		\hat{U} = \prod_{i=1}^{N} \exp \left(-\rmi \tilde{\theta}_i \hat{J}_i^y\right) .
	\end{eqnarray}
	Note that $\hat{U}$ indicates a rotation by angle $\tilde{\theta}_i$ along $y$ axis due to
	\begin{eqnarray}
		\exp\left(\rmi \tilde{\theta} \hat{J}_y\right) \hat{J}_x \exp \left(-\rmi \tilde{\theta} \hat{J}_y\right) &=& \cos \tilde{\theta} \hat{J}_x + \sin \tilde{\theta} \hat{J}_z , \\
		\exp\left(\rmi \tilde{\theta} \hat{J}_y\right) \hat{J}_z \exp \left(-\rmi \tilde{\theta} \hat{J}_y\right) &=& \cos \tilde{\theta} \hat{J}_z - \sin \tilde{\theta} \hat{J}_x .
	\end{eqnarray}
	The unitary transformation $\hat{U}$ 
	brings the rotated $z$ axis along the direction of the large spin obtained from the mean-field approach \cite{PhysRevB.71.224420}.
	
	Then we can apply the Holstein-Primakoff transformation \cite{PhysRev.58.1098}, which maps the collective spin operators to bosonic creation and annihilation operators as
	\begin{subequations}
		\begin{eqnarray}
			\hat{J}_i^z &=& J - \hat{a}_i^{\dagger} \hat{a}_i,\\
			\hat{J}_i^+ &=& \sqrt{2 J - \hat{a}_i^{\dagger} \hat{a}_i} \hat{a}_i ,\\
			\hat{J}_i^- &=& \hat{a}_i^{\dagger} \sqrt{2 J - \hat{a}_i^{\dagger} \hat{a}_i} ,
		\end{eqnarray}
	\end{subequations}
	with $\left[\hat{a}_i, \hat{a}_j^{\dagger}\right] = \delta_{i, j}$.
	The thermodynamic limit corresponds to $J \rightarrow + \infty$. Due to $J  \gg \expect{\hat{a}_i^{\dagger} \hat{a}_i}$, the Holstein-Primakoff transformation can be approximately simplified as
	\begin{subequations}\label{eq:HP}
		\begin{eqnarray} 
			\hat{J}_i^z &=& J - \hat{a}_i^{\dagger} \hat{a}_i,\\
			\hat{J}_i^+ &\approx& \sqrt{2 J} \hat{a}_i ,\\
			\hat{J}_i^- &\approx& \sqrt{2 J} \hat{a}_i^{\dagger} .
		\end{eqnarray}
	\end{subequations}
	Substituting Eq. (\ref{eq:HP}) into Eq. (\ref{eq:H_R}), we obtain a low-energy effective Hamiltonian as follows,
	\begin{eqnarray}
		\hat{\bar{H}} &\approx& J E +  J^{1/2} \hat{\bar{H}}_1 + J^{0} \hat{\bar{H}}_2, \label{eq:H_eff}\\
		\hat{\bar{H}}_1 &=& - \frac{1}{\sqrt{2}} \sum_{i=1}^N \left( \sin \tilde{\theta}_i + \lambda \cos \tilde{\theta}_{i} \left(\sin \tilde{\theta}_{i + 1} + \sin \tilde{\theta}_{i - 1}\right) \right) \nonumber\\
		&& \left(\hat{a}_i^{\dagger} + \hat{a}_i\right) , \label{eq:H1}\\
		\hat{\bar{H}}_2 &=& \sum_i^N \bar{\epsilon}_i \hat{a}_i^{\dagger} \hat{a}_i + \frac{\bar{\chi}_{i, i+1}}{2} \left(\hat{a}_i^{\dagger} + \hat{a}_i\right) \left(\hat{a}_{i + 1}^{\dagger} + \hat{a}_{i + 1}\right) \label{eq:H2}
	\end{eqnarray}
	with 
	\begin{subequations} \label{eq:para_eff}
	\begin{eqnarray}
		\bar{\epsilon}_i &=&  \cos \tilde{\theta}_i - \lambda \sin \tilde{\theta}_{i} \left(\sin \tilde{\theta}_{i - 1} + \sin \tilde{\theta}_{i + 1}\right), \\
		\bar{\chi}_{i,j} &=& \lambda \cos \tilde{\theta}_i \cos \tilde{\theta}_{j} = \bar{\chi}_{j,i},
	\end{eqnarray}
	\end{subequations}
	where we have ignored terms of order $J^{-l}$ with $l > 0$.
	The first term in Eq. (\ref{eq:H_eff}) corresponds to the energy expectation value (\ref{eq:E_GS}) given by the mean-field approach. Minimizing the energy expectation value gives rise to the ground-state energy, which also eliminates $\hat{\bar{H}}_1$, as indicated by Eqs. (\ref{eq:E_min}) and (\ref{eq:H1}). Finally, we only need to deal with the quadratic Hamiltonian $\hat{\bar{H}}_2$ which can be solved exactly by the symplectic transformation \cite{Serafini2017-sk}.
	
	Before performing the symplectic transformation, we first introduce the vector of canonical operators $\hat{\mathbf{r}} = \left(\hat{x}_1, \hat{x}_2, \hat{x}_3, \hat{p}_1, \hat{p}_2, \hat{p}_3\right)^T$ with  $\hat{x}_i = \left(\hat{a}_i^{\dagger} + \hat{a}_i\right) / \sqrt{2}$ and $\hat{p}_i = \rmi \left(\hat{a}_i^{\dagger} - \hat{a}_i\right) / \sqrt{2}$. $\hat{\mathbf{r}}$ satisfies the canonical commutation relation $\left[\hat{\mathbf{r}}, \hat{\mathbf{r}}^T\right] = \rmi \Omega$, where $\Omega$ is given by
	\begin{eqnarray}
		\Omega = \left(\begin{array}{rr}
			O_3 & I_3 \\
			-I_3 & O_3
		\end{array}\right) ,
	\end{eqnarray}
	with $O_3$ and $I_3$ the $3 \times 3$ null and  identity matrices respectively.
	In terms of the vector of canonical operators $\hat{\mathbf{r}}$, the quadratic Hamiltonian $\hat{\bar{H}}_2$ can be rewritten as
	\begin{eqnarray}
		\hat{\bar{H}}_2 &=& \sum_{i=1}^3 \frac{\bar{\epsilon}_i}{2} \left(\hat{x}_i^2 + \hat{p}_i^2 - 1\right) + \bar{\chi}_{i, i+1} \hat{x}_i \hat{x}_{i + 1} \\
		&=& \frac{1}{2} \hat{\mathbf{r}}^T H \hat{\mathbf{r}} - \sum_{i=1}^3 \frac{\bar{\epsilon}_i}{2},  \nonumber
	\end{eqnarray}
	with the Hamiltonian matrix 
	\begin{eqnarray}
		H = H_x \oplus H_p
	\end{eqnarray}
	and
	\begin{equation}
		H_x = \left(\begin{array}{ccc}
			\bar{\epsilon}_1 & \bar{\chi}_{1, 2} & \bar{\chi}_{1,3} \\
			\bar{\chi}_{2,1} & \bar{\epsilon}_2 & \bar{\chi}_{2,3} \\
			\bar{\chi}_{3,1} & \bar{\chi}_{3,2} & \bar{\epsilon}_3
		\end{array}\right), \quad
		H_p = \left(\begin{array}{ccc}
			\bar{\epsilon}_1 & 0 & 0 \\
			0 & \bar{\epsilon}_2 & 0 \\
			0 & 0 & \bar{\epsilon}_3
		\end{array}\right) .
	\end{equation}

\begin{figure}
		\includegraphics[scale=0.9]{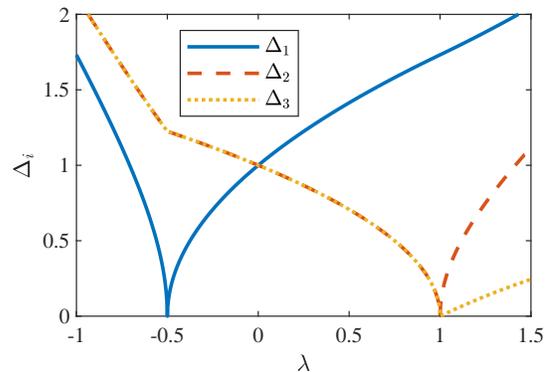}
		\caption{\label{fig:exc} The excitation energy $\Delta_i$ as a function of the dimensionless coupling strength $\lambda$. The blue solid line, red dashed line and yellow dotted line correspond to $\Delta_1$, $\Delta_2$ and $\Delta_3$ respectively.}
	\end{figure}

	According to Williamson's theorem \cite{Serafini2017-sk}, for the positive defined real matrix $H$, there exists a symplectic transformation $S$ ($S^{T} \Omega S = \Omega$) such that 
	\begin{equation}
		S^{T} H S = D, \text{ with } D = \text{diag} \left(\Delta_1, \Delta_2, \Delta_3, \Delta_1, \Delta_2, \Delta_3\right) .
	\end{equation}
	One can introduce a new vector of canonical operators $\hat{\mathbf{r}}' = S^{-1} \hat{\mathbf{r}}$ which split the quadratic Hamiltonian into decoupled degrees of freedom as
	\begin{eqnarray}
		\hat{\bar{H}}_2 &=& \sum_{i=1}^3 \frac{\Delta_i}{2} \left(\hat{x}_i^{\prime 2} + \hat{p}_i^{\prime 2}\right) - \sum_{i=1}^3 \frac{\bar{\epsilon}_i}{2}  \\
		&=& \frac{1}{2} \hat{\mathbf{r}}'^T D \hat{\mathbf{r}}' - \sum_{i=1}^3 \frac{\bar{\epsilon}_i}{2} . \nonumber
	\end{eqnarray}
	$\Delta_i \ge 0$ corresponds to the excitation energy, as shown in Fig. \ref{fig:exc}. Particular attention should be paid to the lowest excitation energies, namely, $\Delta_{\text{min}} = \min \left(\Delta_1, \Delta_2, \Delta_3\right)$, as it corresponds to the energy gap between the ground state and the first excited state. Clearly, the energy gap closes ($\Delta_{\text{min}} \rightarrow 0$) for  $\lambda \rightarrow \lambda_{\pm}$, which is a characteristic signature of the quantum phase transition.
	
\begin{figure}
		\includegraphics[scale=0.94]{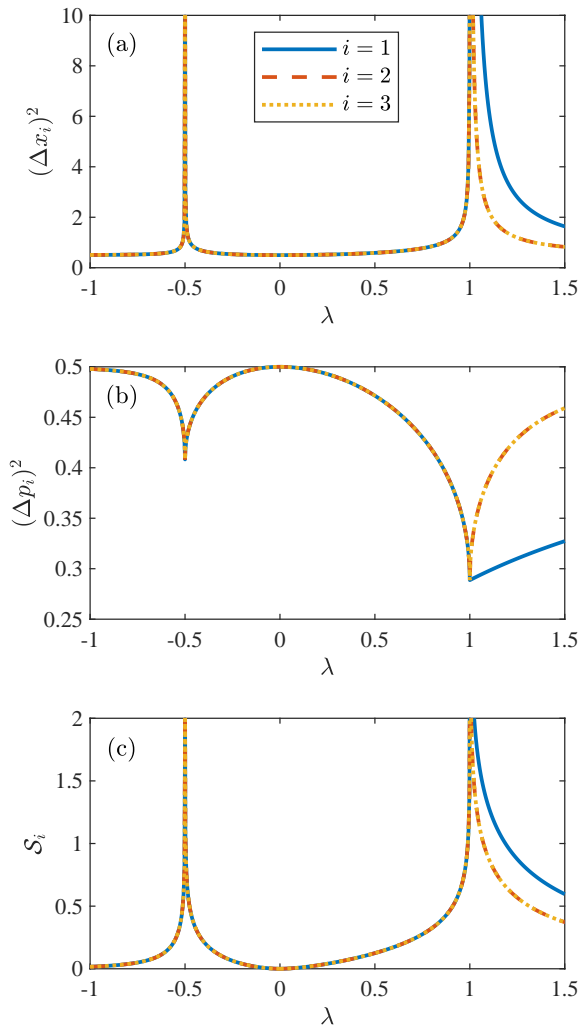}
		\caption{\label{fig:dx_dp_se} (a) The quantum fluctuation in $x_i$ quadrature $\left(\Delta x_i\right)^2$, (b) the quantum fluctuation in $p_i$ quadrature $\left(\Delta p_i\right)^2$ and (c)  the von Neumann entropy $\mathcal{S}_i$ as a function of the dimensionless coupling strength $\lambda$ for the $i$th large spins. The blue solid line, red dashed line and yellow dotted line correspond to  site index $i=1,2,3$ respectively.}
	\end{figure}

	It is well-known that the ground state of the quadratic Hamiltonian is a Gaussian state, which plays a significant role in continuous variable quantum information \cite{Serafini2017-sk}.
	Instead of dealing with the infinite dimension of the associated Hilbert space, one only needs to deal with the $6 \times 6$ covariance matrix $\sigma$ which is capable to give the complete description of an arbitrary Gaussian state (up to local unitary operations) \cite{Serafini2017-sk,Adesso_2007}. The covariance matrix $\mathbf{\sigma}$ is defined as
	\begin{eqnarray}
		\mathbf{\sigma} = \frac{1}{2} \expect{\left\{\left(\hat{\mathbf{r}} - \expect{\hat{\mathbf{r}}}\right), \left(\hat{\mathbf{r}} - \expect{\hat{\mathbf{r}}}\right)^T\right\}} ,
	\end{eqnarray}
	with $\expect{\hat{A}}$ corresponding to the expectation value of operator $\hat{A}$. Given the symplectic transformation $S$, it's easy to confirm that the covariance matrix can be written as $\mathbf{\sigma} = S S^T / 2$ \cite{Adesso_2007}. The quantum fluctuations in $\hat{x}_i$ and $\hat{p}_i$ are characterized by the standard deviations which are given by the diagonal elements of the covariance matrix $\sigma$, namely,
	\begin{subequations}\label{eq:fluctuation}
	\begin{eqnarray}
		\left(\Delta x_i\right)^2 &=& \expect{\hat{x}_i^2} - \expect{\hat{x}_i}^2 = \sigma_{i,i}, \\
		\left(\Delta p_i\right)^2 &=& \expect{\hat{p}_i^2} - \expect{\hat{p}_i}^2 = \sigma_{3+i, 3+i} .
	\end{eqnarray}
	\end{subequations}
	The quantum fluctuations in $x_i$ and $p_i$ quadrature are shown in Figs. \ref{fig:dx_dp_se}(a) and \ref{fig:dx_dp_se}(b) respectively. Near the quantum critical point $\lambda_{\pm}$, the quantum fluctuation in $x_i$ quadrature tends to exponentially diverge, whereas that in $p_i$ quadrature tends to a finite value. Furthermore, $\left(\Delta p_i\right)^2 < \frac{1}{2}$ indicates a strong squeezing effect especially near the quantum critical point \cite{scully_zubairy_1997}.
	
	The von Neumann entropy, which characterizes the entanglement between different subsystems, is directly related to the Heisenberg's uncertainty relation for the quadratic Hamiltonian of interacting bosonic systems \cite{PhysRevA.86.043807,Adesso_2007}.
	In terms of $\Delta x_i$ and $\Delta p_i$, the von Neumann entropy $\mathcal{S}_i$ can be written as
	\begin{eqnarray} \label{eq:se}
		\mathcal{S}_i &=& \left(\Delta x_i \Delta p_i + \frac{1}{2}\right) \log \left(\Delta x_i \Delta p_i + \frac{1}{2}\right) \nonumber\\
		&& - \left(\Delta x_i \Delta p_i - \frac{1}{2}\right) \log \left(\Delta x_i \Delta p_i - \frac{1}{2}\right) ,
	\end{eqnarray}
	which  describes the entanglement between the $i$th large spin and the other two large spins. A growing interest has recently been devoted to the study of quantum phase transitions from the entanglement point of view. As illustrated in Fig. \ref{fig:dx_dp_se} (c), the von Neumann entropy diverges near the quantum critical point $\lambda_{\pm}$, which indicates the strong correlations between different large spins ignored by the mean-field approach.
	
	From now on, we present the detailed analytical exact results for the excitation energy, quantum fluctuation, and von Neumann entropy, especially near the quantum critical point in different phases.
	
	\begin{figure*}
		\includegraphics[scale=0.83]{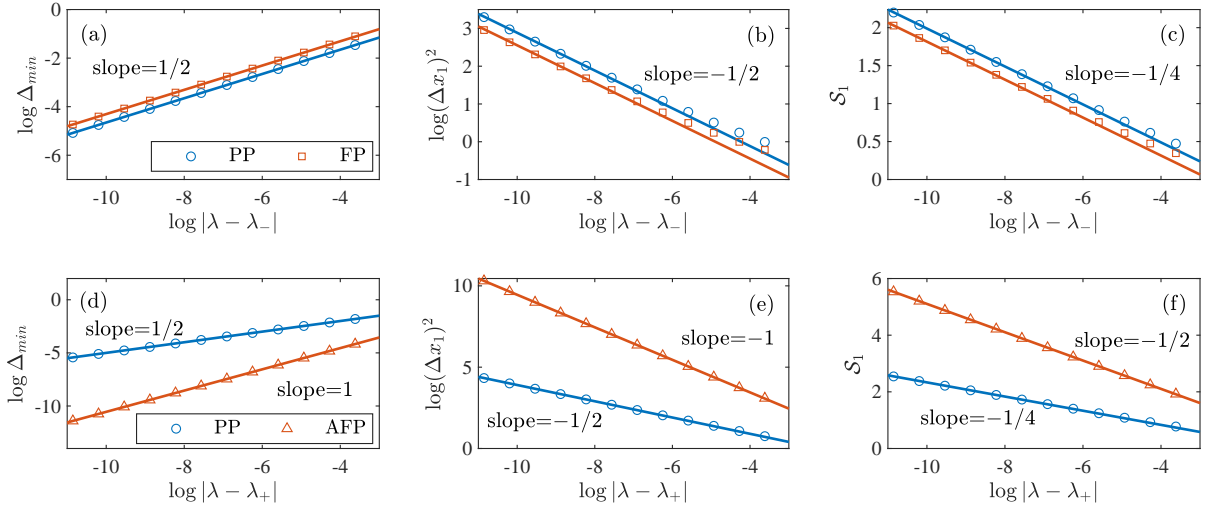}
		\caption{\label{fig:critical} Critical behaviors near the quantum phase transition point. (a) and (d) show the lowest excitation energy $\Delta_{\text{min}}$ as a function of the coupling strength $\lambda$ near $\lambda_-$ and $\lambda_{+}$ respectively. (b) and (e) show the quantum fluctuation in the $x_1$ quadrature $\left(\Delta x_1\right)^2$. (c) and (f) show the von Neumann entropy $\mathcal{S}_1$. The circle, square, and triangle correspond to the analytical exact results obtained from the symplectic transformation for the paramagnetic phase (PP), ferromagnetic phase (FP), and antiferromagnetic phase (AFP) respectively. The solid lines refer to the linear fitted results whose slopes are shown near each line. }
	\end{figure*}
	
	\subsection{Disordered paramagnetic phase}
	
	In the disordered paramagnetic phase ($\lambda_- < \lambda < \lambda_+$), $\tilde{\theta}_i = 0$ leads to $\bar{\epsilon}_i = 1$ and $\bar{\chi}_{i,j} = \lambda$ [Eq. (\ref{eq:para_eff})]. The quadratic Hamiltonian corresponds to a translationally invariant harmonic chain with nearest neighbor interactions. According to the Bloch's theorem, the wavenumber in the first Brillouin zone corresponds to $k = 0$ and $\pm 2 \pi / 3$. The symplectic transformation can be obtained analytically \cite{Serafini2017-sk}, with which we achieve the excitation energies $\Delta_1 = \sqrt{1 + 2 \lambda}$ for $k=0$ and $\Delta_2 = \Delta_3 = \sqrt{1 - \lambda}$ for $k=\pm 2 \pi / 3$. 
	The lowest excitation energy corresponds to $\Delta_{\text{min}} = \Delta_1$ for $\lambda_- < \lambda < 0$ and $\Delta_{\text{min}} = \Delta_2 = \Delta_3$ for $0 < \lambda < \lambda_+$. Therefore, the critical behavior associated with the excitation energy is $\Delta_{\text{min}} \propto \left(\lambda - \lambda_-\right)^{1/2}$ for $\lambda \rightarrow \lambda_-$, whereas it is $\Delta_{\text{min}} \propto \left(\lambda_+ - \lambda\right)^{1/2}$ for $\lambda \rightarrow \lambda_+$, as shown in Figs. \ref{fig:critical}(a) and \ref{fig:critical}(d). 
	
	The quantum fluctuations obtained from the symplectic transformation can be written as
	\begin{subequations}
	\begin{eqnarray}
		\left(\Delta x_i\right)^2 &=& \frac{1}{6} \left[2 \left(1 - \lambda\right)^{-1/2} + \left(1 + 2 \lambda\right)^{-1/2}\right], \\
		\left(\Delta p_i\right)^2 &=& \frac{1}{6} \left[2 \left(1 - \lambda\right)^{1/2} + \left(1 + 2 \lambda\right)^{1/2}\right] .
	\end{eqnarray}
	\end{subequations}
	Near the quantum critical  point which corresponds to the paramagnetic-ferromagnetic phase transition ($\lambda \rightarrow \lambda_-$), the quantum fluctuation in $p_i$ quadrature tends to a finite value with $\left(\Delta p_i\right)^2 \rightarrow \sqrt{6} / 6$, whereas the exponentially diverged quantum fluctuation in $x_i$ quadrature is given by
	\begin{eqnarray} \label{eq:dx_-}
		\left(\Delta x_i\right)^2 \propto \left( \lambda - \lambda_-\right)^{-1/2} ,
	\end{eqnarray}
	as shown in Fig. \ref{fig:critical} (b).
	Near the quantum critical  point which corresponds to the paramagnetic-antiferromagnetic phase transition ($\lambda \rightarrow \lambda_+$), the quantum fluctuation in $p_i$ quadrature tends to a finite value with $\left(\Delta p_i\right)^2 \rightarrow \sqrt{3} / 6$, whereas the exponentially diverged quantum fluctuation in $x_i$ quadrature is given by
	\begin{eqnarray} \label{eq:dx_+}
		\left(\Delta x_i\right)^2 \propto \left( \lambda_+ - \lambda\right)^{-1/2}, 
	\end{eqnarray}
	as shown in Fig. \ref{fig:critical} (e).
	
	The quantum fluctuation in $x_i$ quadrature given by Eqs. (\ref{eq:dx_-}) and (\ref{eq:dx_+}) can be unified as
	\begin{eqnarray} \label{eq:dx_order}
		\left(\Delta x_i\right)^2 \propto \left| \lambda - \lambda_{\pm}\right|^{-1/2} .
	\end{eqnarray}
	Since $\Delta x_i \Delta p_i \gg 1$ near the quantum critical point, we can simplify the von Neumann entropy in Eq. (\ref{eq:se}) as
	\begin{eqnarray} \label{eq:se_disorder}
		\mathcal{S}_i \approx \log \Delta x_i \Delta p_i \approx -\frac{1}{4} \log \left|\lambda - \lambda_{\pm}\right| + C_i,
	\end{eqnarray}
	with $C_i$ a constant independent of the coupling strength,
	which explains the divergence in $\mathcal{S}_i$ as shown in Figs. \ref{fig:critical}(c) and \ref{fig:critical}(f).
	
	\subsection{Ordered phase}
	
	\subsubsection{Ferromagnetic phase}
	In the ferromagnetic phase ($\lambda<\lambda_-$), the global $Z_2$ symmetry is broken.
	From Eqs. (\ref{eq:theta_ferro}) and (\ref{eq:para_eff}), we obtain 
	\begin{subequations} \label{eq:para_eff_ferro}
		\begin{eqnarray}
			\bar{\epsilon}_i &=& \bar{\epsilon} = - 2  \lambda, \\
			\bar{\chi}_{i,j} &=& \bar{\chi} = \frac{1}{4 \lambda} .
		\end{eqnarray}
	\end{subequations}
	Therefore, $\bar{\epsilon}_i$ and $\bar{\chi}_{i,j}$ don't depend on the site index, similar to those in the paramagnetic phase. The excitation energies correspond to $\Delta_1 = \sqrt{4 \lambda^2 - 1}$ and $\Delta_2 = \Delta_3 = \sqrt{4 \lambda^2 + \frac{1}{2}}$. Near the quantum critical point ($\lambda \rightarrow \lambda_-$), the lowest excitation energy satisfies $\Delta_{\text{min}} = \Delta_1 \propto \left(\lambda_- - \lambda\right)^{1/2}$, as shown in Fig. \ref{fig:critical} (a).
	The quantum fluctuations are given by
	\begin{subequations}
	\begin{eqnarray}
		\left(\Delta x_i\right)^2 &=& \frac{1}{6} \left[2 \left(1 - \frac{\bar{\chi}}{\bar{\epsilon}}\right)^{-1/2} + \left(1 + 2 \frac{\bar{\chi}}{\bar{\epsilon}}\right)^{-1/2}\right], \\
		\left(\Delta p_i\right)^2 &=& \frac{1}{6} \left[2 \left(1 - \frac{\bar{\chi}}{\bar{\epsilon}}\right)^{1/2} + \left(1 + 2 \frac{\bar{\chi}}{\bar{\epsilon}}\right)^{1/2}\right] .
	\end{eqnarray}
	\end{subequations}
	Similar to those in the paramagnetic phase, near the quantum critical  point ($\lambda \rightarrow \lambda_-$), the quantum fluctuation in $p_i$ quadrature tends to a finite value with $\left(\Delta p_i\right)^2 \rightarrow \sqrt{6} / 6$, whereas the exponentially diverged quantum fluctuation in $x_i$ quadrature is given by
	\begin{eqnarray}
		\left(\Delta x_i\right)^2 &\propto& \left(\lambda_- - \lambda\right)^{-1/2} ,
	\end{eqnarray}
	as shown in Fig. \ref{fig:critical} (b). Accordingly, the von Neumann entropy shares the same form as that in Eq. (\ref{eq:se_disorder}).
	
	\subsection{Frustrated antiferromagnetic  phase}
	
	In the frustrated antiferromagnetic phase ($\lambda > \lambda_+$), both the global $Z_2$ symmetry and the translational symmetry are broken. $\tilde{\theta}_i$ depends on the site index $i$, so do other properties related with $\tilde{\theta}_i$. The exact analytical expressions for $\Delta_i$, $\left(\Delta x_i\right)^2$, $\left(\Delta p_i\right)^2$ and $\mathcal{S}_i$ become much more tedious. Therefore, we just give the critical behaviors near the quantum critical point ($\lambda \rightarrow \lambda_+$). The lowest excitation energy satisfies $\Delta_{\text{min}} \propto \left(\lambda - \lambda_+\right)$, as verified in Fig. \ref{fig:critical} (d). Near the quantum critical  point which corresponds to the paramagnetic-antiferromagnetic phase transition ($\lambda \rightarrow \lambda_+$), the quantum fluctuation in $p_i$ quadrature tends to a finite value with $\left(\Delta p_i\right)^2 \rightarrow \sqrt{3} / 6$, whereas the exponentially diverged quantum fluctuation in $x_i$ quadrature is given by
	\begin{eqnarray}
		\left(\Delta x_i\right)^2 &\propto&  \left(\lambda - \lambda_+\right)^{-1} ,
	\end{eqnarray}
	as shown in Fig. \ref{fig:critical} (e).
	Accordingly, the von Neumann entropy near the quantum critical point depicted in Fig. \ref{fig:critical} (f)  can be written as
	\begin{eqnarray} \label{eq:se_AFP}
		\mathcal{S}_i \approx \log \Delta x_i \Delta p_i \approx -\frac{1}{2} \log \left(\lambda - \lambda_{+}\right) + C'_i.
	\end{eqnarray}

	Obviously, the critical behaviors of the excitation energy, quantum fluctuation in $x_i$ quadrature, and von Neumann entropy in the frustrated antiferromagnetic phase are different from those in both paramagnetic and ferromagnetic phases, which demonstrate the significance of the geometric frustration. A mnemonic summary of the critical behaviors in different quantum phases is provided in Table  \ref{tab:summary}.
	
	It should be noted that the aforementioned analyses are valid in the thermodynamic limit ($J \rightarrow +\infty$). Beyond the thermodynamic limit (finite $J$), higher order terms proportional to $J^{-l}$ with $l>0$ are not negligible after the Holstein-Primakoff transformation, which leads to a nonquadratic effective Hamiltonian. Such a Hamiltonian cannot be solved by the symplectic transformation, but rather more sophisticated techniques, such as the continuous unitary transformation \cite{PhysRevB.71.224420,https://doi.org/10.1002/andp.19945060203,PhysRevD.48.5863,PhysRevLett.93.237204}. Neither the global $Z_2$ symmetry nor the translational symmetry will be broken for finite $J$, which can be confirmed by the absence of the nonzero order parameter $J_i^x$.  The higher-order terms 
	can be employed to study the finite-size scaling behavior near the quantum critical point \cite{PhysRevB.71.224420,PhysRevLett.93.237204}, which is beyond the scope of this work.
	
	\begin{table*}
		\caption{\label{tab:summary}
			Excitation energy $\Delta_{\text{min}}$, quantum fluctuation $\left(\Delta x_i\right)^2$ and von Neumann entropy $\mathcal{S}_i$ near the critical point.}
		\begin{ruledtabular}
			\begin{tabular}{cccc}
				&Ferromagnetic phase ($\lambda < \lambda_-$)&Paramagnetic phase ($\lambda_- < \lambda < \lambda_+$)&Antiferromagnetic phase ($\lambda > \lambda_+$)\\ \hline
				$\Delta_{\text{min}}$&$\left(\lambda_- - \lambda\right)^{1/2}$
				&$\left|\lambda - \lambda_{\pm}\right|^{1/2}$\footnote[1]{$\left(\lambda - \lambda_-\right)^{1/2}$ for $\lambda \rightarrow \lambda_-$ and $\left(\lambda_+ - \lambda\right)^{1/2}$ for $\lambda \rightarrow \lambda_+$}&$\left(\lambda - \lambda_+\right)$\\
				$\left(\Delta x_i\right)^2$&$\left(\lambda_- - \lambda\right)^{-1/2}$&$\left|\lambda - \lambda_{\pm}\right|^{-1/2}$\footnote[2]{$\left(\lambda - \lambda_-\right)^{-1/2}$ for $\lambda \rightarrow \lambda_-$ and $\left(\lambda_+ - \lambda\right)^{-1/2}$ for $\lambda \rightarrow \lambda_+$}
				&$\left(\lambda - \lambda_+\right)^{-1}$\\
				$\mathcal{S}_i$&$-\frac{1}{4} \log \left(\lambda_- - \lambda\right)$&$-\frac{1}{4} \log \left|\lambda - \lambda_{\pm}\right|$\footnote[3]{$-\frac{1}{4} \log \left(\lambda - \lambda_-\right)$ for $\lambda \rightarrow \lambda_-$ and $-\frac{1}{4} \log \left(\lambda_+ - \lambda\right)$ for $\lambda \rightarrow \lambda_+$}&$-\frac{1}{2} \log \left(\lambda - \lambda_+\right)$
			\end{tabular}
		\end{ruledtabular}
	\end{table*}

	\section{Conclusions} \label{sec:summary}

	As a natural generalization of the TFIM, the coupled-top model describes large spins with ferromagnetic or antiferromagnetic interaction. It has been widely employed to study the quantum phase transition,  chaotic phenomena and quantum scar, etc. In this paper, we study three interacting large spins located on a triangle. Compared with the original coupled-top model with two large spins, the third large spin cannot simultaneously satisfy antiferromagnetic interaction with the other two, which introduces the geometric frustration. Similar phenomena have been found in the Rabi ring \cite{PhysRevLett.127.063602,PhysRevLett.129.183602}, which can be mapped into a frustrated magnetic system with interacting large spins. Here we focus on the quantum phase transition in the triangular coupled-top model, especially on the novel critical behaviors induced by the geometric frustration.
	
	The triangular coupled-top model admits three phases: disordered paramagnetic phase, ferromagnetic phase, and frustrated antiferromagnetic phase, depending on whether the ground state preserves or breaks the global $Z_2$ symmetry and translational symmetry. The quantum critical points which separate three phases, as well as the order parameters $J_i^x$ in different phases, can be obtained by the mean-field approach. In the paramagnetic phase ($\lambda_-< \lambda < \lambda_{+}$), the ground state is non-degenerate. Both the global $Z_2$ symmetry and translational symmetry are preserved with $J_i^x = 0$. In the ferromagnetic phase ($\lambda < \lambda_-$), only the translational symmetry is preserved and the ground state is two-fold degenerate. The breaking of the global $Z_2$ symmetry leads to nonzero order parameters $J_i^x$. In the antiferromagnetic phase ($\lambda > \lambda_+$) where the geometric frustration comes into play, the ground state is six-fold degenerate. Neither the global $Z_2$ symmetry nor the translational symmetry is preserved. The order parameter $J_i^x$ depends on the site index $i$.
	
	Further insight into the quantum effects beyond the mean-field approach can be achieved by employing the Holstein-Primakoff transformation and the symplectic transformation.  After the Holstein-Primakoff transformation, three interacting large spins are mapped into coupled harmonic oscillators described by the quadratic Hamiltonian, which can be solved exactly by the symplectic transformation. The energy gap between the ground state and first excited state closes near the quantum critical point. The quantum fluctuation in $x_i$ quadrature tends to exponentially diverge, whereas that  in $p_i$ quadrature tends to be a finite value, which is a signature of the squeezing effect. The von Neumann entropy also diverges near the critical point, which indicates a strong correlations between different large spins. It is noteworthy that the critical behaviors in the antiferromagnetic phase are different from those in the paramagnetic and ferromagnetic phases.
	The triangular coupled-top model opens up new opportunities to investigate novel critical behaviors induced by the geometric frustration.

	\begin{figure}
		\includegraphics[scale=0.45]{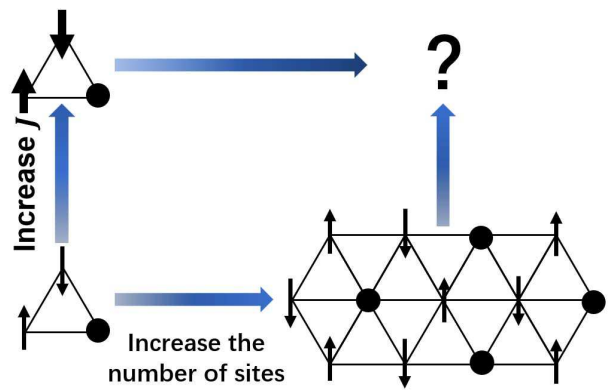}
		\caption{\label{fig:triangle} Bottom-left corner corresponds to an elementary unit formed by three spin-$\frac{1}{2}$s ($J=\frac{1}{2}$) on the triangle. Increasing the number of sites leads to the TFIM on a triangular lattice, as shown in the bottom-right corner. Increasing $J$ on each site leads to the triangular coupled-top model in this work, as shown in the top-left corner. The thin and thick arrows correspond to the spin-$\frac{1}{2}$ and large spin respectively. The dot refers to the frustrated spin.}
	\end{figure}

	The geometric frustration plays a significant role in both the triangular coupled-top model and the TFIM on a triangular lattice with antiferromagnetic interaction. They can be regarded as a generalization of an elementary unit formed by three spin-$\frac{1}{2}$s ($J=\frac{1}{2}$) on the triangle, as shown in Fig. \ref{fig:triangle}. 
	The elementary unit has no quantum phase transition. The Ising model on a triangular lattice is achieved by increasing the number of sites, which manifests much richer phenomena.
	In the absence of the transverse field, it exhibits a macroscopic degeneracy and is disordered at $T=0$ \cite{PhysRev.79.357}. Introducing the transverse field leads to the TFIM, which possesses a three-sublattice state with long-range magnetic order due to the mechanism known as order by disorder \cite{PhysRevB.63.224401}. 
	The coupled-top model is achieved by replacing the spin-$\frac{1}{2}$s in the elementary unit with large spins ($J \gg \frac{1}{2}$), which yields novel critical behavior in the frustrated antiferromagnetic phase. It is interesting to 
	replace spin-$\frac{1}{2}$s of the TFIM on a triangular lattice by large spins,
	which are left to future research.

	\begin{acknowledgments}
		L. D. is supported by Zhejiang Provincial Natural Science Foundation of China under Grant No. LQ23A050003. Y.-Z. W. is supported by the National Natural Science Foundation of China under Grant No. 12105001 and Natural Science Foundation of Anhui Province under Grant No. 2108085QA24. Q.-H. C. is supported by the National Science Foundation of China under Grant No. 11834005.
	\end{acknowledgments}
	
	%
	
\end{document}